\documentclass[onecolumn,superscriptaddress,aps,floatfix,12pt]{revtex4-1}
\usepackage{hyperref}
\usepackage{blindtext}
\usepackage{amsmath}
\usepackage{amssymb}
\usepackage{graphicx}
\usepackage{mathrsfs}
\usepackage{braket}
\usepackage{bm,bbm}
\usepackage{color}
\usepackage{multirow}

\begin{document}

\title{Exciton polariton critical non-Hermitian skin effect with spin-momentum-locked gains}
\author{Xingran Xu}
\affiliation{School of Science, Jiangnan University, Wuxi 214122, China}
\author{Lingyu Tian}
\affiliation{Beijing Academy of Quantum Information Sciences, Beijing, 100193, P.R. China}

\author{Zhiyuan An}
\affiliation{Beijing Academy of Quantum Information Sciences, Beijing, 100193, P.R. China}

\author{Qihua Xiong}
\email{qihua{\_}xiong@tsinghua.edu.cn}
\affiliation{State Key Laboratory of Low-Dimensional Quantum Physics and Department of Physics, Tsinghua University, Beijing, 100084, P.R. China}
\affiliation{Beijing Academy of Quantum Information Sciences, Beijing, 100193, P.R. China}
\affiliation{Frontier Science Center for Quantum Information, Beijing, 100084, P.R. China}
\affiliation{Collaborative Innovation Center of Quantum Matter, Beijing 100084, P.R. China}

\author{Sanjib Ghosh}
\email{sanjibghosh@baqis.ac.cn}
\affiliation{Beijing Academy of Quantum Information Sciences, Beijing, 100193, P.R. China}

\date{\today}

\begin{abstract}
The critical skin effect, an intriguing phenomenon in non-Hermitian systems, displays sensitivity to system size and manifests distinct dynamical behaviors. In this work, we propose a novel scheme to achieve the critical non-Hermitian skin effect of exciton polaritons in an elongated microcavity system. We show that by utilising longitudinal-transverse spin splitting and spin-momentum-locked gain, a critical non-Hermitian skin effect can be achieved in a continuous system without the need of an underlying lattice. We find that a phase transition can be induced by changing the cavity detuning with respect to the exciton energy. We identify a measurable order parameter associated with this phase transition and demonstrate the corresponding critical behavior. Our work offers a flexible approach to manipulate non-Hermitian phases of exciton polaritons, thereby expanding the potential applications of polaritonic devices.
\end{abstract}
\maketitle

\section{Introduction}
Recently, there has been a surge of interest in non-Hermitian systems, characterized by their complex energy spectra, in contrast to the real-valued spectra observed in Hermitian systems~\cite{PhysRevLett.80.5243,Bender_2007}. Non-Hermitian systems show many intriguing properties such as non-conservation of the particle number~\cite{Rev1,Rev2}, parity-time symmetry induced exceptional points ~\cite{zdemir_2019,R_ter_2010}, and non-reciprocal transport~\cite{ Fruchart_2021, Wakatsuki_2017}. Among many fascinating features of non-Hermitian systems, one particularly intriguing phenomenon is known as the non-Hermitian skin effect (NHSE)~\cite{El_Ganainy_2018, Linfop_2023, Zhang_rev2022, Ding_2022}. In this system, all states are localized at the edge, with a complex energy spectrum that is highly sensitive to the boundary conditions~\cite{Topoorg,song2019, Kawabata2019}. While translation symmetry is crucial in the Bloch theorem for typical condensed matter systems, it dramatically breaks down in non-Hermitian systems. A systematic method has been developed to define a unified Brillouin Zone, bridging the theory of Hermitian and non-Hermitian systems and extending the bulk-boundary correspondence~\cite{Shelykh_2009, Jiangping2020, Yokomizo_2020, Xiao_2020}. In this theory, the Bloch factor  $|e^{ik}|\neq 1$ and the effective momentum is considered a complex number. Here, a well-defined topological invariant is evaluated from the energy spectra under periodic boundary conditions (PBC) and open boundary conditions (OBC)~\cite{jiang2018, Kawabata_2019,yao2018}. NHSE has been experimentally realized and studied in diverse platforms, such as electric circuits~\cite{Helbig_2020,PhysRevResearch.2.023265,PhysRevResearch.2.022062}, cold atoms~\cite{coldNHSE1,coldNHSE2}, acoustic systems ~\cite{zhang2021acoustic,PhysRevB.106.134112}, and photonic systems ~\cite{Zhu2020,PhysRevApplied.14.064076}.

Intriguingly, when two systems with non-Hermitian skin effects, each with oppositely localized sides, are coupled, a novel critical behavior emerges known as the critical non-Hermitian skin effect (CNHSE)~\cite{Li_2020}.
In contrast to NHSE, CNHSE describes a unique critical behavior, characterized by discontinuous jumps in the energy spectrum and wave functions across a critical point~\cite{Li_2020,Liu2020}. In CNHSE, wave function localization is size-sensitive, displaying distinct dynamic behaviors across different phases of the system. Moreover, critical skin states exhibit certain scale-free behaviors, decaying exponentially in space, contrasting with conventional skin states typically associated with power-law spatial decay~\cite{Linfop_2023, Scaling1,Scaling2}.

Here, we consider exciton polaritons in microcavity systems~\cite{Rev1,Rev2} as a platform to realise CNHSE. As hybrid quasi-particles exciton polaritons arise from the strong coupling between excitons and photons within semiconductor microcavities~\cite{ghosh2022microcavity}. Exciton polaritons exhibit rich physical effects such as TE-TM splitting~\cite{Dufferwiel2015,Kavokin2015}, spin-momentum locking~\cite{Spin-Momentu1,Spin-Momentu2}, and ultrafast dynamics~\cite{chen2023unraveling,xu2023ultrafast}. Additionally, microcavities exhibits highly non-Hermitian behaviours due to the finite quality factor (Q factor)~\cite{Basov_2016,Paik_2022}. Various methods have been proposed to realise NHSE in polariton systems. For instance, under a coherent optical pump, fluctuations on top of polariton condensates shift towards the edges, resulting in NHSE with a momentum shift in the bandstructures~\cite{Xu2021}. A variety of artificial lattices are utilised to achieve different form of NHSE such as zigzag chains ~\cite{Mandal2020prl}, ring lattices~\cite{huawen2021},  elliptical micropillars~\cite{mandal2021topological} and in two-dimensional lattices ~\cite{XU2022}. While achieving NHSE is intriguing, realizing CNHSE can provide greater control over the non-Hermitian dynamics of polaritons and lead to uncovering novel physical effects. Experimentally, CNHSE is typically observed in macroscopic classical systems like electric circuits~\cite{PhysRevResearch.2.022062}. In contrast, exciton polaritons can undergo Bose-Einstein condensation with long-range phase coherence, enabling the occurrence of CNHSE in the quantum domain, which could be harnessed for future quantum technologies~\cite{o2009photonic}.

\section{Microcavity polaritons} 
\begin{figure}[t]
\centering
\includegraphics[width=\columnwidth]{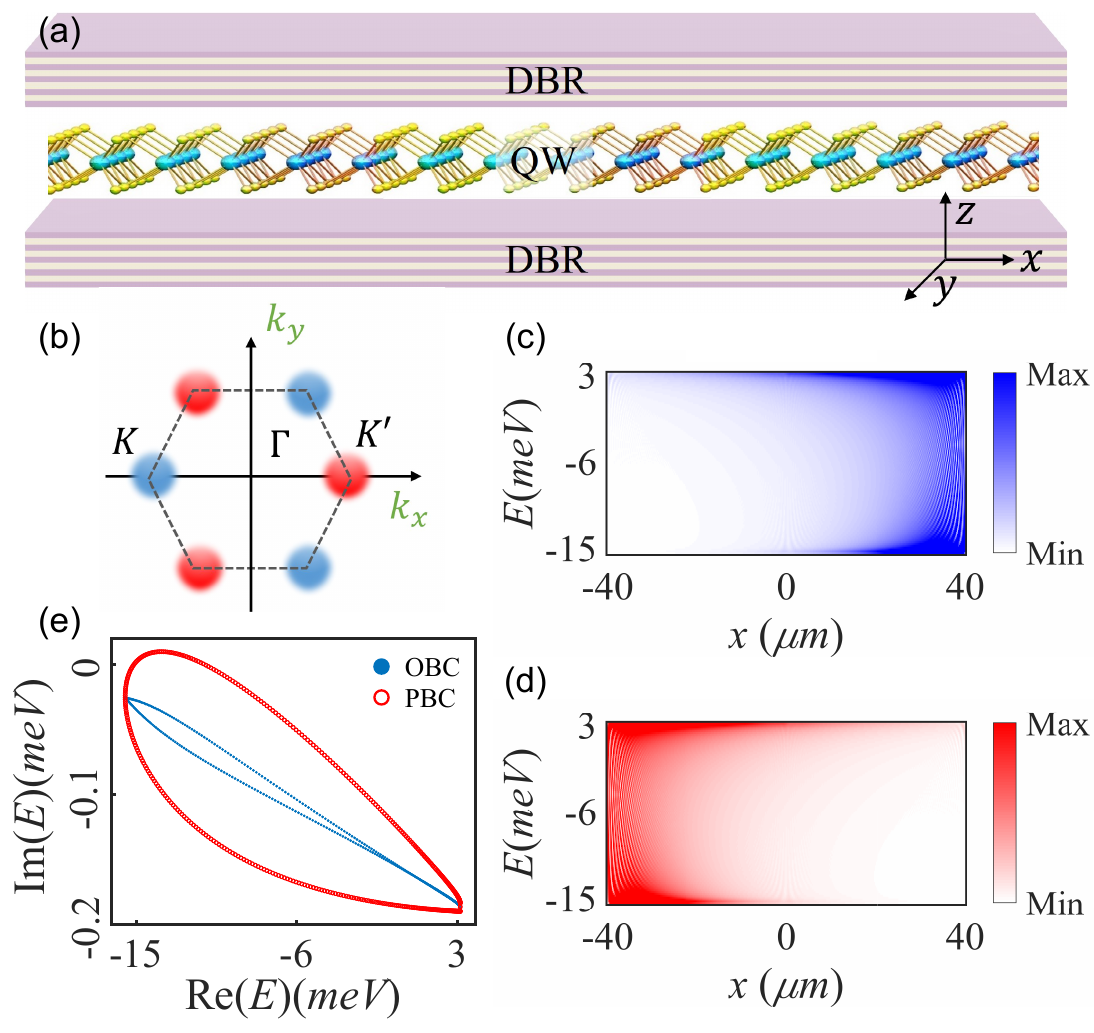}
\caption{Microcavity exciton polaritons with valley-spin-locked gain. (a) An effective one-dimensional microcavity comprises a quantum well positioned between a pair of DBR mirrors. (b) Visualization of valley-spin-locked emission in 1D reciprocal space. (c, d) Spin-resolved local density of states and (e) complex energy spectrum illustrating the $\mathbb{Z}_2$ non-Hermitian skin effect. Parameters are: $\gamma=0.2 meV$  and $\Omega=40 meV$ for (c)-(d).}
\label{model}
\end{figure}

Here, our emphasis is on the lower polariton branch of exciton-polaritons, characterized by the following dispersion relation: $E_{LP}(\bm{k}) = [ E_p(k)+E_x-\sqrt{\Omega^2+(E_p(k)-E_x)^2}]/2$, where $E_p(\bm{k})=\hbar^2k^2/(2m_p)$ is the cavity photon dispersion, $E_x$ is the exciton energy relative to the photon energy (cavity detuning), and $\Omega$ is the Rabi energy splitting~\cite{ghosh2022microcavity}. Experimentally, one can tune the relative energy detuning $E_x$ between the exciton and cavity photons, and the effective photon mass $m_p$ by choosing the appropriate distance between the DBR mirrors. Here, we consider  $m_p=10^{-4} m_e$ with $m_e$ is the bare electron's mass. Furthermore, here we consider one-dimensional (1D) systems, which can be effectively obtained by confining polaritons in the transverse directions of microcavities, as shown in Fig.~\ref{model}(a).

Owing to their hybrid nature, exciton polaritons can exhibit unique properties by utilizing novel quantum well materials or specific microcavity configurations. Intriguingly, quantum wells also serve as optical gain materials, generating new particles upon external excitation. This enables the design of exotic gain profiles by leveraging the properties of the quantum wells. In this context, we envision quantum wells with robust spin-momentum-locking, where exciton reservoirs can serve as the incoherent spin-momentum-locked gain for the lower polariton branch. This scenario leads to an effective Hamiltonian,
\begin{equation}
H_\sigma= E_{LP}(\hat{k})+if(\sigma, k),
\end{equation}
where $f(\sigma, k)$ represents the spin-momentum-locked gain, and $\sigma$ is the spin index.

For instance, in monolayer transition metal dichalcogenides (TMDs), the spin-momentum-locking phenomenon manifests as the valley-spin-locking in the reciprocal space. Under nonresonant excitation, excitons at the $K$ and $K'$ valleys are generated with opposite spins. Consequently, the optical emissions resulting from exciton recombination show opposite polarization in the $K$ and $K'$ valleys~\cite{onga2017exciton}. Similarly, spin-valley-locked emission can be achieved using perovskite metasurfaces~\cite{chen2023compact}. These systems can act as valley-spin-locked gain media for microcavity exciton polaritons, as depicted in Fig.~\ref{model}(b). We envision that such gain media can be characterized by an effective momentum-dependent gain function:
\begin{equation}
f(\sigma, k)= \gamma \exp[ {-\hbar^2({\bm{k}} - \tau k_0 )^2/\Delta_k^2}], \label{H12}
\end{equation}
where the valley index $\tau=\pm 1$ is locked to the spin index $\sigma$~\cite{XiaoDi2012}. The phenomenological parameters $\gamma$, $k_0$, and $\Delta_k$ represent the strength, position, and width of the effective gain in reciprocal space. We consider $k_0=3\,\mu m^{-1}$ and $\Delta_k=4.5\,\mu m$. The valley-spin-locked gain in Eq.~(\ref{H12}) pulls polaritons towards either left or right end of the system depending on their polarization. This underlying mechanism induces the non-Hermitian $\mathbb{Z}_2$ skin effect~\cite{Topoorg}. For practical observation, we consider the spin-resolved local density of states:
\begin{equation}
\rho_\sigma(x,E) = -\text{Im}[\langle x | \hat{G}^\sigma_R(E) |x \rangle]
\end{equation}
where $\hat{G}^\sigma_R(E) = \lim_{\eta \to 0}[E-H_\sigma+i\eta]^{-1}$ is the retarded Green's operator, and $\ket{x}$ is the state corresponding to the position $x$. $\rho_\sigma(x,E)$ quantifies the probability of finding polaritons at position $x$ with energy $E$ and polarization $\sigma$. Physically, the local density of states can be estimated with space and energy-resolved measurements~\cite{Su_2021}. In Fig.~\ref{model}(c) and (d), we show $\rho_\sigma(x,E)$  for two polarization components $\sigma=\pm 1$. We find that indeed $\rho_\sigma(x, E)$ is localized on the left (for $\sigma=+1$) or right (for $\sigma=-1$) end of the system for all energy $E$. This is analogous to non-Hermitian skin effects found in lattice systems~\cite{PhysRevResearch.2.022062,zhang2021universal,Kawabata20202d}. Furthermore, in Fig.~\ref{model}(e), we show the real and imaginary parts of the energies $E$ corresponding to the eigenmodes for open and periodic boundary conditions. We find clear signatures of point group topology associated with our system. We emphasize that here, the non-Hermitian skin effect exists in a continuous system without the need for an underlying lattice.

\begin{figure}[h]
\centering
\includegraphics[width=\columnwidth]{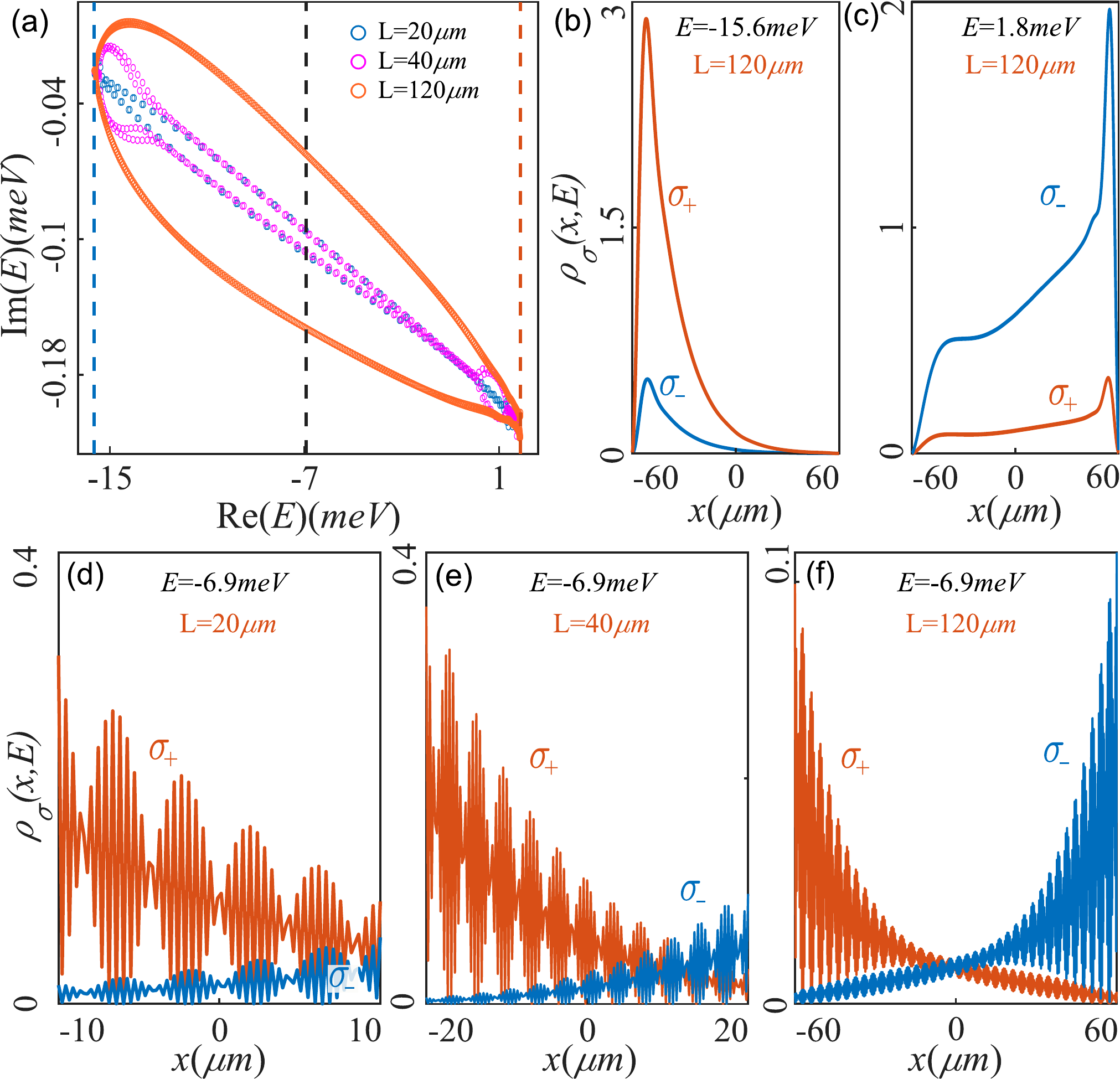}
\caption{Non-Hermitian critical skin effect of microcavity exciton-polaritons. (a) The complex spectrum for increasing system size $L=20,~40$ and $120\,\mu m$. (b, and c) are local density of states at the two ends of the spectrum $E=-15.6\, meV$ (dotted blue vertical line in (a)) and $1.8\, meV$ (dotted red vertical line in (a)). (d, e, and f) are the local density of states at the middle of the spectrum (dotted black vertical line in (a)) for three different system sizes. Parameters are: $R=0.02 \,meV$, $\gamma=0.2 \,meV$  and $\Omega=40 \,meV$.}
\label{OBC}
\end{figure}

 \section{Critical non-Hermitian skin effect}
 Degenerate exciton polariton modes with opposite circular polarization can experience both longitudinal-transverse splitting ~\cite{leyder2007observation}  and Zeeman splitting induced by an external magnetic field~\cite{polimeno2021tuning,lyons2022giant, PhysRevB.100.121303}. In our effective 1D system, these spin splittings can be expressed using a two-component Hamiltonian:
\begin{equation}
 H =\left[ {\begin{array}{cc}H_++V_0-i\Gamma & R \\R & H_- -V_0 -i\Gamma \\\end{array} } \right] \label{poskin}
\end{equation}
where $R\simeq \Delta_{LT} (k_0)$ is the longitudinal-transverse (TE-TM) splitting around the wave vector $k_0$~\cite{andreani1988longitudinal,kavokin2005optical}, $V_0$ is the Zeeman splitting, and $\Gamma$ is a constant decay.  $H_\pm$ is the Hamiltonian for different spin described by Eq. (\ref{H12}). Here, we consider $\gamma-\Gamma=10^{-2} meV$ where $\hbar/\Gamma$ is the life time of the polariton.

In Fig.~\ref{OBC}(a), we plot complex spectrum of the Hamiltonian in Eq.~(\ref{poskin}) for different system size $L$. For small $L$, spectrum show small splitting between two curves. For larger $L$, we observe the formation of a closed loop in the complex spectrum. This is one of the important features of CNHSE where a discontinuous change occur in the spectrum under OBC~\cite{Kawabata2023cri,Li_2020}.
A corresponding topological invariant can be defined: $W=[2\pi i]^{-1}\int d\log\left[\det(H-E_R)\right]$, where $E_R$ is the complex eigenvalues with OBC and $H$ is the Hamiltonian with PBC~\cite{Topoorg}.

The signatures of CNHSE can also be seen in the local density of states $\rho_\sigma(x,E)$ at different energies and system sizes. As shown in Fig.~\ref{OBC}(b) and (c), $\rho_\sigma(x,E)$ near edges of the spectrum, for instance at energy $1.8\,meV$ ($-15.6\,meV$), tends to localize at the left (right) edge of the system for both spins $\sigma =\pm 1$. In contrast, $\rho_\sigma(x,E)$ near the middle of the spectrum, such as at energy $-6.9\,meV$, has a transition from one-sided localization (without symmetry) to two-sided localization (with symmetry) when $L$ is increased from $L=20\,\mu m$ to $120\,\mu m$, as shown in Figs.~\ref{OBC}(d)-(f). This clear demonstration of system-size sensitivity of the skin effect confirms CNHSE in our system~\cite{Li_2020}.

\begin{figure}[b]
\centering
\includegraphics[width=1.0\columnwidth]{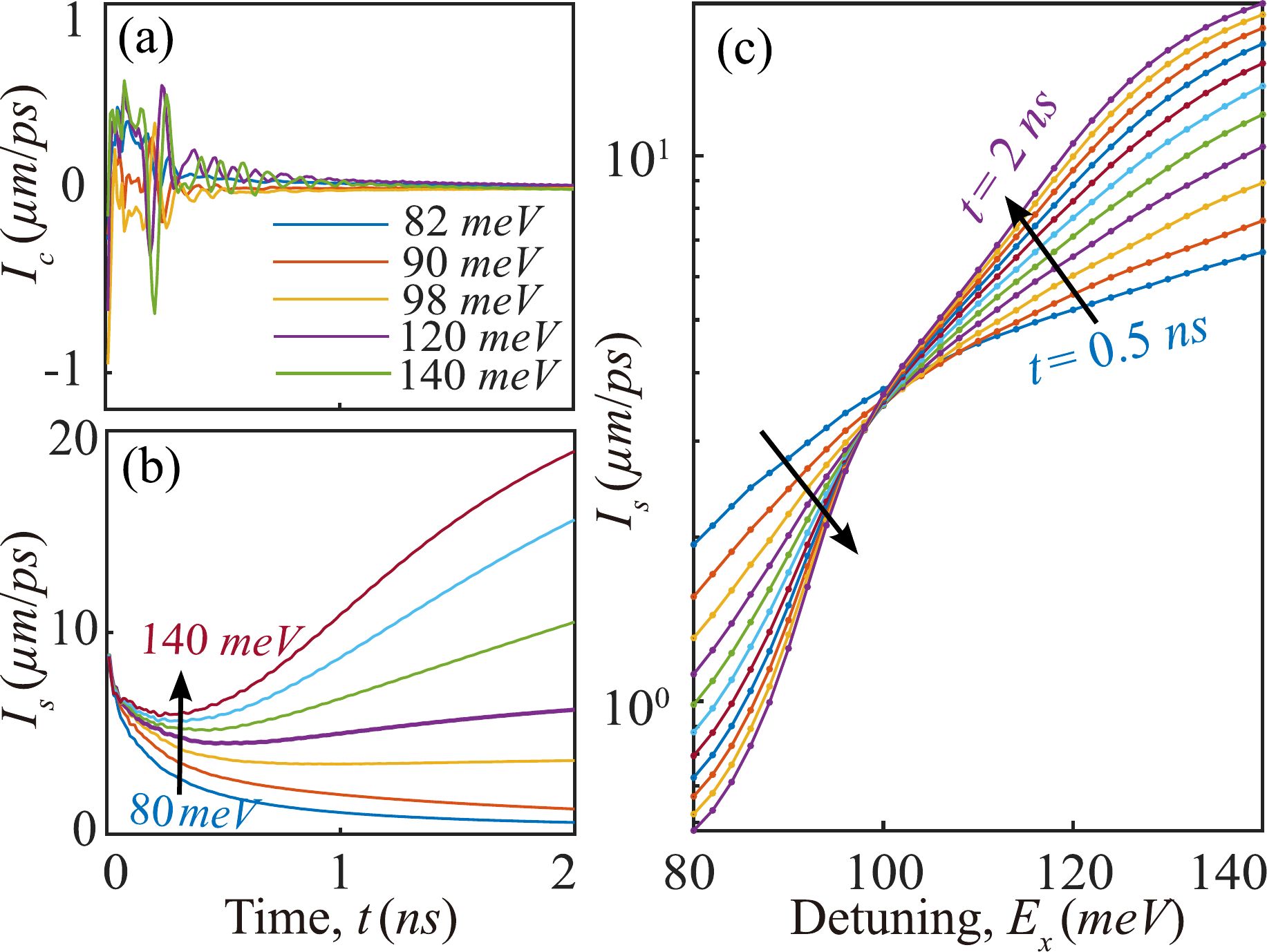}
\caption{Phase transition of microcavity exciton polaritons with spin-momentum-locked gain. The time evolution of the average mass currents (a) and spin currents (b). The spin currents of different times as a function of the exciton detuning (c). All curves intersect at a critical point $E_x \approx 98 \, meV$, which is the phase transition point. The direction of arrow represents an increase in detuning or time. We consider the following parameters: $\gamma=0.2\, meV$, $R=0.2\, meV$, $V_0=0.02\,meV$, $L= 28 \,\mu m$,  and $\Omega=40 \, meV$.}
\label{transition}
\end{figure}

\section{Phase transition and order parameter}
Here, we show a novel phase transition in semiconductor microcavities induced by CNHSE of polaritons. We find that this phase transition can be fully controlled by the detuning between the cavity photons and exciton energy. As a measurable quantity we explore transport properties of polaritons. In a closed quantum system at thermodynamic equilibrium the total current flow is zero. However, in open systems non-zero currents can be induced by skin effects ~\cite{Kawabata2023cri}. For our system, we define spin resolved polariton currents as,
\begin{equation}
I_\sigma (t) =\frac{\hbar}{\rho(t) m_\text{eff} } \int \text{Re}[ \psi_\sigma^* (x,t) \partial_x \psi_\sigma(x,t) ] dx
\end{equation}
where $\partial_x\equiv \partial/\partial x$, and $\rho (t) = \sum_\sigma\int |\psi_\sigma(x,t) |^2 dx $ is the total density of polaritons. The current is normalised by the density $\rho (t)$ to avoid the occurrence of divergence due the non-Hermitian nature of the system. Furthermore, to study the transport of polariton mass and spin separately, we define time-average mass current ($I_c(t)$) and spin current ($ I_s (t)$):
\begin{eqnarray}
 I_c(t) = \frac{1}{t} \int_0^t \sum_\sigma  I_\sigma(t') dt', \notag\\
 I_s (t) = \frac{1}{t} \int_0^t  \sum_\sigma \sigma I_\sigma(t') dt'.
\end{eqnarray}

In our system, we consider that initially unpolarized polaritons are injected with a Gaussian profile $\psi_\pm(x,t=0)= A e^{-x^2/\sigma_x^2}$ which is centered at the middle of the system and $\sigma_x$ ($\sigma_x=0.01\,\mu m$) is the width of the profile. We find that the average mass current rises at short times, but vanishes $I_c (t\to \infty ) = 0 $ in the long time limit because of the reciprocity~\cite{Kawabata2023cri}, as shown in Fig.~\ref{transition}(a). However, as shown in Figs.~\ref{transition}(b) we find that the average spin current $I_s$ has intriguing features and can be controlled by changing the energy detuning $E_x$ between the excitons and the cavity photons.

We observe a critical energy threshold $E_c \approx 98\, meV$, where the long-time spin current $I_s$ tends towards zero for $E_x < E_c$, while it approaches a nonzero value for $E_x > E_c$. This behavior indicates a phase transition, where the two phases are characterized by the long-time spin current $I_s(t\to \infty)$. This becomes more evident when plotting $I_s(t)$ as a function of $E_x$ for different times, as shown in Fig.~\ref{transition}(c). We observe that as time progresses, $I_s(t)$ evolves towards a step-like function, indicating the presence of two phases around $E_x = E_c$. This observed feature of $I_s(t)$ implies that it is a critical observable quantity that undergoes a transition from zero to a nonzero value across the phase transition in the long time regime. Moreover, all curves in Fig.~\ref{transition}(c) intersect near the critical point $E_x \approx 98\,meV$, a behavior commonly observed at a phase transition point~\cite{ghosh2015coherent,Singh20}. We emphasize that such a phase transition is possible due to the intriguing shape of the polariton dispersion, which allows the effective polariton mass to be tuned by changing the cavity detuning.

For numerical calculations, it is crucial to note that the total particle number is not conserved during the time evolution because of the non-Hermitian nature of the Hamiltonian. To avoid any divergence in the physical quantities, one can use normalised particles numbers.

\section{Phase diagram}
\begin{figure}[t]
\centering
\includegraphics[width=1\columnwidth]{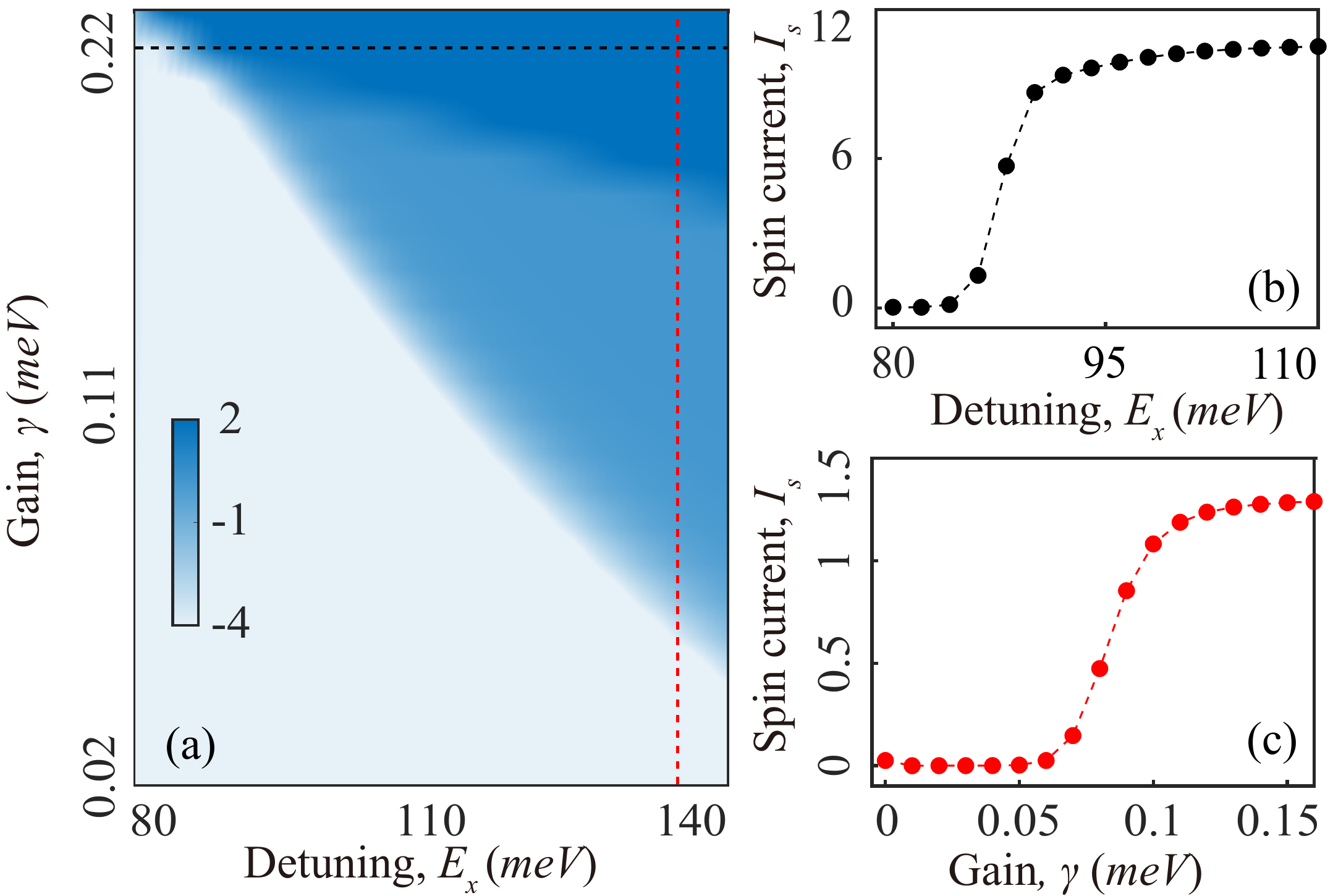}
\caption{Phase diagram of microcavity exciton polaritons with spin-momentum-locked gain. (a) The logarithmic of the steady-state spin current ($\ln I_{s}$) as a function of $E_x$ and $\gamma$. We find a distinct boundary separating the two phases characterized by the value of $I_{s}$. (b) Spin current as a function of cavity detuning $E_x$ at fixed gain $\gamma=0.22\, meV$. (c) Spin current as a function $\gamma$ for a fixed $E_x=135\, meV$. We consider the following parameters: $R=0.2\, meV$, $V_0=0.02\,meV$, $L= 28 \mu m$,  and $\Omega =40\, meV$.}
\label{pd}
\end{figure}

We find that the spin current can be controlled by both the strength of the spin-momentum-locked gain $\gamma$ and the exciton detuning $E_x$. Fig.~\ref{pd}(a) depicts a phase diagram for microcavity exciton polaritons, where distinct phases (characterized by different values of $I_s(t\to \infty)$) are delineated by a color code for different values of the gain $\gamma$ and the detuning $E_x$. As shown in Fig.~\ref{pd}(a), two distinct phases are clearly discernible, distinguished by different shades of colors. For $E_x<0$, the system stays in an equilibrium phase with zero spin current $I_s(t\to\infty)=0$. As $E_x$ increases, the spin current rises from zero to nonzero, leading the system to reach a non-equilibrium state. The momentum-locked gain $\gamma$ serves as the source of non-Hermiticity, inducing the spin current and controlling the phase transition. As illustrated in Figs.~\ref{pd}(b) and (c), the phase transition can be induced by adjusting either the parameter $\gamma$ while keeping $E_x$ constant or by varying $E_x$ while maintaining $\gamma$ constant.

For physical observation of the phase transition, open cavity systems can be considered, where adjusting the exciton detuning energy $E_x$ in semiconductor microcavities can be achieved by altering the distance between the top and bottom mirrors~\cite{lackner2021tunable}. Furthermore, we note that the transition can also be realized by tuning $\gamma$, which can be achieved by controlling the power of excitation to the quantum well. However, it is important to keep in mind that the parameter $\gamma$ is an effective parameter, which might not have a monotonic relation with the external power.

\section{Discussion}
We developed a scheme to realise the critical skin effect in semiconductor microcavities by utilizing spin-momentum-locked gains. In contrast to the existing approaches, our method operates on a continuous model, eliminating the requirement for underlying lattice structures. We have shown that the complex energy spectrum of exciton polaritons under CNHSE is strongly sensitive to the system size and the local density of states for different spins have different localization behaviors. Moreover, we have demonstrated that a critical non-Hermitian phase transition of polaritons can be induced by changing the detuning between the cavity photons and exciton energy. We have shown that the phase transition can be fully characterized by the average spin current.

Experimentally, spin-momentum-locked gains can be obtained in TMD microcavities through the valley-spin-locking mechanism. This mechanism is facilitated by the unique electronic band structures arising from strong spin-orbit coupling and inversion symmetry breaking. Moreover, microcavity exciton polaritons have the capability to undergo Bose-Einstein condensation, thereby enabling the critical skin effect to extend into the quantum realm. This opens up possibilities for the development of future polaritonic devices with greater control over spin degrees of freedom.

\section{Acknowledgements}
This work is supported by the Excellent Young Scientists Fund Program (Overseas), the National Natural Science Foundation of China (Grant No. 12274034), and the Fundamental Research Funds for the Central Universities (Grant No. JUSRP123027). Q. X. gratefully acknowledges support from the National Natural Science Foundation of China (Grant No. 12020101003, 92250301, and 12250710126) and the State Key Laboratory of Low-Dimensional Quantum Physics at Tsinghua University.

\bibliographystyle{apsrev4-1}
\bibliography{my}

\end{document}